\documentclass[a4paper,twocolumn,aps,prb,showpacs]{revtex4}
\usepackage{amsmath}
\usepackage{bbm}
\usepackage{amsfonts}
\usepackage{amssymb}
\usepackage{graphicx}
\usepackage{hyperref}

\begin{document}

\author{Heng \surname{Wang}}
\author{Guido \surname{Burkard}}
\affiliation{Department of Physics, University of Konstanz, D-78457 Konstanz, Germany}

\title{Mechanically induced spin resonance in a carbon nanotube}
\begin{abstract}
The electron spin is a promising qubit candidate for quantum computation and quantum information. Here we propose and analyze a mechanically-induced single electron spin resonance, which amounts to a rotation of the spin about the $x$-axis in a suspended carbon nanotube. The effect is based on the coupling between the spin and the mechanical degree of freedom due to the intrinsic curvature-induced spin-orbit coupling. A rotation about the $z$-axis is obtained by the off-resonant external electric driving field. Arbitrary-angle rotations of the single electron spin about any axis in the $x$-$z$ plane can be obtained with a single operation by varying the frequency and the strength of the external electric driving field. With multiple steps combining the rotations about the $x$- and $z$-axes, arbitrary-angle rotations about arbitrary axes can be constructed, which implies that any single-qubit gate of the electron spin qubit can be performed. We simulate the system numerically using a master equation with realistic 
parameters.
\end{abstract}

\maketitle

\section{Introduction}

The characteristics of suspended carbon nanotubes (CNTs)--low mass, high and widely tunable resonance frequencies, and high quality factors \cite{Sazonova2004,Huttel2009,Chaste2011,Laird2012}--make them very promising for nanomechanical devices, e.g., as ultra-sensitive magnetometers, as well as mass and force detectors \cite{Lassagne2008,Chiu2008,Lassagne2011,MoserJ.2013}. The strong coupling between single-electron tunneling and  the mechanical vibration of the suspended CNT \cite{Lassagne2009,Steele2009,Huttel2010,TraversoZiani2011,Ganzhorn2012,Benyamini2014} can be used to probe the CNT's vibration frequency \cite{Garcia-Sanchez2007,Bennett2012}, and the average charge in the quantum dot (QD) on the CNT \cite{Meerwaldt2012}. A suspended CNT can be driven with radio frequency fields into its nonlinear vibrational regime \cite{Huttel2009,Castellanos-Gomez2012}. In a recent theory work where the ground state and the first nonlinear vibration modes of suspended CNTs are used as long lived quantum bits, a two-qubit entangling gate has been proposed by coupling the qubits to an optical cavity \cite{Rips2013}.

QDs in CNTs with a semiconducting band gap have attracted much attention because of the additional valley degree of freedom, among other reasons \cite{Minot2004,P'alyi2011,Pei2012}. 
The spin-orbit interaction was initially expected to be weak in CNTs, but the curvature-induced spin-orbit coupling was later recognized to be 
significant \cite{Entin2001,Chico2004,Min2006,Huertas-Hernando2006,Chico2009,Jeong2009} and to lead to a lifting of the four-fold spin and valley degeneracy, observed in experiment \cite{Kuemmeth2008,Kuemmeth2010}. 
Nevertheless, due to the low nuclear spin density in the carbon-based host material, electron spins in CNT QDs can 
be viewed as prospective quantum bits. 
The spin relaxation in CNT QDs is caused by the spin-orbit coupling and has been investigated in a number  of studies\cite{Bulaev2008,Weiss2010,Jespersen2011}. 
Recently, a spin-orbit coupling as large as $3~ {\rm meV}$ has been measured in a CNT \cite{Steele2013}. 
It was predicted that the inherent spin-orbit coupling can induce a single spin-phonon coupling in a suspended \cite{Rudner2010,Flensberg2010}, CNT which provides a new way for realizing spin-based nanomechanical systems \cite{P'alyi2012,Ohm2012a,Li2012}. 
\begin{figure}[ht!]
	\centering
		{\includegraphics[width=0.45\textwidth]{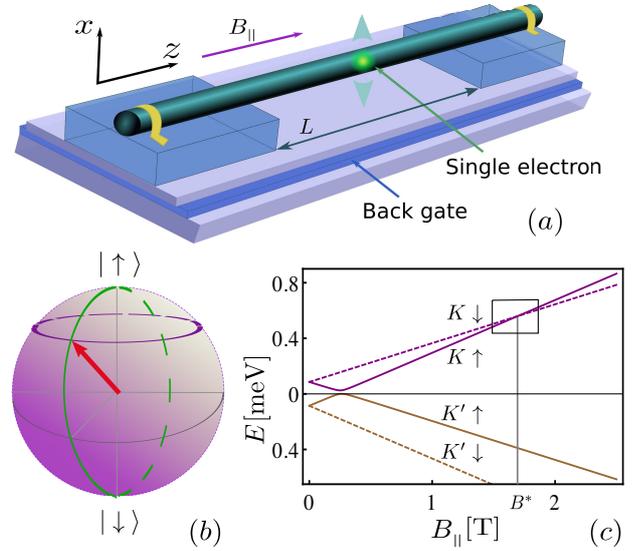}}
\caption{(a) Proposed system for mechanical spin manipulation. A single electron is trapped in a quantum dot (QD) in a suspended carbon nanotube (CNT).  A magnetic field $B_{\parallel}$ parallel to the $z$-axis is applied. A gate voltage applied on the back gate can adjust the resonance frequency of the CNT. An external antenna or the back gate can apply an external ac electric driving field to the charged nanotube. (b) The Bloch sphere representing the quantum state of the electron spin. The green line denotes electron spin resonance (ESR) and the purple line shows a rotation about the $z$-axis. (c)  Energy spectrum of the QD on the CNT as a function of the parallel magnetic field  $B_{\parallel}$. The two degrees of freedom of the valley ($K$, $K'$) and single-electron spin ($\uparrow, \downarrow$) yield a 4-fold degeneracy, which is lifted by $B_{\parallel}$. A pure spin qubit is found around the magnetic field $B^{*}=\Delta_{so}/ (g_s \mu_B)=1.7 \ {\rm T} $. Parameters used here are $ \Delta_{so}=170 \  {\rm \mu eV} $, $\Delta_{KK'}= 12.5 \ {\rm \mu eV} $, $\mu _{orb}=330 \ {\rm \mu eV}$ \cite{P'alyi2012}.
\label{fig:setup}}
\end{figure}
The read-out of single electron spins has 
been proposed by different methods such as making use of magnetic resonance force microscopy \cite{Rugar2004,Rabl2009}, spin blockade \cite{Hanson2005} and spin-phonon coupling \cite{Struck2014}.
In this paper, we propose a mechanically-driven arbitrary manipulation of the electron spin based on this spin-phonon coupling.

 In our proposed nanomechanical system, a suspended CNT is doubly clamped (See Fig. \ref{fig:setup} (a)) \cite{P'alyi2012}. Two electrodes at both ends are used for forming a QD, i.e., trapping of a single electron which can be spin-polarized by an external longitudinal magnetic field $B_{\parallel}$ applied along the $z$-axis of the CNT.   
As we will show, any rotation of the spins about any axis on the $x$-$z$  plane on the Bloch sphere can be obtained with a single-step operation by varying the strength and the frequency of the external driving electric field. Mechanically induced electron spin resonance (ESR) can be performed based on the spin-phonon coupling in the suspended CNT (see Fig. \ref{fig:setup}(b)). The rotation of the spin about the $z$-axis on the Bloch sphere can be implemented by the off-resonant ac electric driving field. Any arbitrary-angle rotation about any arbitrary axis in the whole Bloch sphere can be obtained by a multiple-step operation combining ESR and the rotation of the single spin about the $z$-axis. 

This paper is organized as follows. We first introduce the quantum mechanical system and give the basic Hamiltonian in Sec.~\ref{sec:model}. To explain the mechanism of the mechanically induced ESR, we derive the effective Hamiltonian within a low-energy subspace by using a Schrieffer-Wolff transformation and obtain the respective time-evolution operator in Sec.~\ref{sec:ESR}.  In Sec.~\ref{sec:numerics}, we numerically simulate the time evolutions at different temperatures with realistic parameters. In Sec.~\ref{sec:analysis}, we analyze and discuss the physical mechanism responsible for mechanical ESR. 
We conclude in Sec.~\ref{sec:conclusions} by summarizing the main results.

\section{Model}
\label{sec:model}
To describe the mechanically induced ESR, we model the spin of an electron located in the QD on the CNT, the relevant mechanical resonator mode, as well as the coupling between these two degrees of freedom.
The total Hamiltonian is given by \cite{P'alyi2012}: 
\begin{equation}
\begin{split}
&H=H_0 +H_1 ,\\
&  H_0 = \frac{\hbar \omega_{q}}{2}\sigma_{z}+\hbar \omega_{p}a^{\dag}a,\\
& H_1= 2 \hbar \lambda(a+a^{\dag}) \cos(\omega t)+\hbar g(a+a^{\dag})(\sigma_{+}+\sigma_{-}),
\end{split}
\label{eq:primitive_Hamiltonian}
\end{equation}
where $\sigma_{z}$ is the Pauli matrix describing the electron spin, $\sigma_{\pm}=\sigma_{x}\pm i\sigma_{y}$  are the spin raising and the lowering operators, and $a (a^\dagger)$ is the phonon annihilation (creation) operator.  We assume $\hbar=1$ for simplicity in the following.  We define the two states of the qubit at the crossing point of the two spins in the upper valley (see Fig. \ref{fig:setup}(c)) \cite{Flensberg2010,Jespersen2011,P'alyi2012}. The frequency of the spin splitting is defined as $\omega_q=(B_{\parallel}-B^{*} ) / (\hbar\mu_{B})$, where $B_{\parallel}$ is the applied magnetic parallel field and $B^{*}=\Delta_{so}/(g_s \mu_{B})$,  where $g_s$ is the electron spin g-factor and $\Delta_{so}$ is spin-orbit coupling strength. 

A gate voltage applied on the back gate can adjust the resonance frequency of the CNT \cite{Huttel2009}. By the capacitive coupling, an antenna or the back gate can apply an external ac electric field to drive the charged nanotube \cite{Sazonova2004,Steele2009}. For simplicity, we assume that only a single polarization of the vibration of the CNT can be excited by the ac driving field and the phonon mode has a frequency $\omega_p$. 
The first term in $H_1$ is the driving term caused by the external ac electrical field with a frequency $\omega$ and a strength $\lambda$.
The second term $g(a+a^{\dag})(\sigma_{+}+\sigma_{-})$ describes the spin-phonon coupling which originates from the inherent curvature-induced spin-orbit coupling \cite{P'alyi2012}. The spin-phonon coupling strength is proportional to the spin-orbit coupling, $g \propto\Delta _{so} l_0$, where $l_0$ is the zero-point motion amplitude of the phonon mode. We find a value of the coupling strength $g/2 \pi= 0.56 \ {\rm MHz}$ with realistic parameters $\Delta_{so}= 370 \ {\rm \mu eV} $, $l_0=2.5 \ {\rm pm}$, and $L=400 \ {\rm nm}$ \cite{P'alyi2012}.

A high $Q$ factor and simultaneously a high resonance frequency are desired for reaching the quantum limit of a mechanical system \cite{O'Connell2010,Poot2012}. A $800 \ {\rm nm}$ long carbon nanotube with the resonance frequency $350 \ {\rm MHz}$ and high $Q$ factor $Q\approx150000$ has been found experimentally \cite{Huttel2009, Steele2009}. Recently, it was reported that a straight CNT  with a length of $250 \ {\rm nm}$ and a quality factor $Q=1200$ is used to realize an ultrahigh $4.2 \ {\rm GHz}$ resonance frequency of the flexural fundamental eigenmode \cite{Chaste2011}. We take the resonance frequency $\omega_p/2 \pi$ as $1.5 \ {\rm GHz}$ and the quality factor as  $Q\approx95000$ corresponding to $\Gamma=\omega_p/Q\approx10^5 \ {\rm s^{-1}}$  in between these two cases.

To simulate the system in a realistic way, we use a master equation in which the external phonon bath and the damping of the phonon mode are taken into account, 

 \begin{equation}
	\begin{split}
	\dot{\rho}=& -\frac{i}{\hbar}[H,\rho] +(n_{B}+1)\Gamma \left(a\rho a^{\dagger} -\frac{1}{2}\{a^{\dagger} a,\rho\}\right) \\
	& +n_{B}\Gamma\left(a^{\dagger}\rho a-\frac{1}{2}\{a a^{\dagger},\rho\}\right),
\end{split}
	\label{eq:master_eqn_finiteT}
\end{equation}
where $n_{B}=1/(e^{\hbar \omega_{p}/k_{B}T}-1)$ is the Bose-Einstein occupation factor of the phonon bath at temperature $T$. We assume a low spontaneous qubit relaxation rate $1/T_1\ll g$ due to the low densities of the other phonon modes near the qubit frequency \cite{Mariani2009, Rudner2010, P'alyi2012}.

\section{Electron spin resonance}
\label{sec:ESR}

At low temperatures, $k_B T \ll \hbar \omega_p $, we can assume for simplicity that there are only four relevant states $| 0 \downarrow \rangle$, $|0 \uparrow  \rangle$, $|1 \downarrow  \rangle$  and $|1 \uparrow \rangle$ in this system. Dividing the Hilbert space into two subspaces with $0$ and $1$ phonons, we find that
$H_{0}$ is block-diagonal  and $H_{1}$ is block-off-diagonal on the space of these four states 
\begin{equation}
H_{0}=\left( \begin{array}{cc}
H_{0}^{(0)}& 0\\ 
0& H_{0}^{(1)}  \end{array} \right) ,
\ H_{1}=\left( \begin{array}{cc}
0& H_{1}^{(01)}\\ 
H_{1}^{(10)}& 0  \end{array} \right) ,
\end{equation}
where $H_{0}^{(i)}(i=0,1)$ operate on the states with $i$ phonons, and $H_{1}^{(ij)} (i,j=0,1,\ i\neq j)$ are the interaction terms between states with $i$ and $j$ phonons. Their concrete forms are $H_0^{(0)}=\frac{\hbar \omega_{q}}{2}\sigma_{z}$, $H_0^{(1)}=\frac{\hbar \omega_{q}}{2}\sigma_{z}+\hbar \omega_{p}$, 
$H_1^{(01)}= 2 \hbar \lambda(a+a^{\dag}) \cos(\omega t)$, and $H_1^{(10)}=\left(H_1^{(01)}\right)^\dagger$.
To understand the time evolution of the spin, it is useful to obtain the effective Hamiltonian of the lowest-energy subspace which contains the states $|0 \downarrow \rangle$ and $|0 \uparrow \rangle$.
For this purpose, a Schrieffer-Wolff transformation can be applied and an effective total Hamiltonian which is block-diagonal is obtained.  We assume that there is a large detuning of the frequency $\Delta=\omega_p-\omega \gg g,\lambda$ between the driving field and the phonon mode, and also between the phonon mode and the qubit mode $\Delta \geq \omega_p-\omega_q  \gg g,\lambda$. Because the external driving term $2 \hbar \lambda(a+a^{\dag}) \cos(\omega t)$ in Eq. (\ref{eq:primitive_Hamiltonian}) is 
time-dependent, a time-dependent Schrieffer-Wolff transformation is applied \cite{Goldin2000}. Assuming a unitary transformation $U(t)$ is a function of time and $\psi$ is the original wave function, the wave function is transformed as $\tilde{\psi}=U{\psi}$. With the Schr\"{o}dinger equation $ i \partial_t \psi=H \psi$ and $ i \partial_t \tilde{\psi}=\tilde{H} \tilde{\psi}$, one can obtain the transformed Hamiltonian after the time-dependent Schrieffer-Wolff transformation as
\begin{equation}
\tilde{H}=UHU^\dag-iU(\partial_t U^\dag).
\label{eq:derive}
\end{equation}
Writing $U(t)=e^{S(t)}$, where $S(t)=-S(t)^\dagger \propto O(H_1)$, we obtain
\begin{equation}
\tilde{H}=e^{S(t)} H e^{-S(t)}+ie^{S(t)}\partial_t e^{-S(t)}.
\label{eq:schrieffer2}
\end{equation}
Expanding Eq.(\ref{eq:schrieffer2}) in a Taylor series in $S(t)$, one can gain the following transformed Hamiltonian $\tilde{H}$ at second order 
\begin{equation}
\begin{split}
\tilde{H}&=H_0+\underbrace{H_{1}+[S(t),H_{0}]+i\dot{S}(t)}_{O(H_1)}+O(H_1^3)\\
& +\underbrace{[S(t),H_{1}]+\frac{1}{2}[S(t),[S(t),H_0]]+\frac{1}{2} i[S(t),\dot{S}(t)]}_{O(H_1^2)}\\
&+O(H_1^3).
\end{split}
\end{equation}
Due to the time dependence, we obtain one additional first order term $i\dot{S}(t)$ and one second order term $\frac{1}{2} i[S(t),\dot{S}(t)]$. To obtain a block-diagonal Hamiltonian, the first order terms $O(H_1)$ which are block-off-diagonal are eliminated by the condition,
\begin{equation}
H_{1}+[S(t),H_{0}]+i\dot{S}(t)=0.
\label{eq:firstorder}
\end{equation}
From Eq. (\ref{eq:firstorder}) we can obtain an expression for $S(t)$ and the effective Hamiltonian $ \tilde{H}$ which only contains the zeroth order and the second order terms in $H_1$. Substituting Eq. (\ref{eq:firstorder}) into Eq. (\ref{eq:zerothandsecond}), we find a simple form of the effective Hamiltonian, 
\begin{equation} 
\begin{split}
\tilde{H}= H_0+\frac{1}{2}[S(t),H_1].
\label{eq:zerothandsecond}
\end{split}
\end{equation}

We now solve Eq.(\ref{eq:firstorder}), a first-order inhomogeneous differential equation for $S(t)$. We obtain the simplest solution for $S(t)$ as 
\begin{equation} 
S(t)= \left( 
\begin{array}{cc}
0 & S_1(t) \\
-S_1^\dagger(t)& 0 
\end{array} \right),
\label{eq:smatrix}
\end{equation}
where
\begin{equation}
\begin{split}
S_1(t)=
\left( 
\begin{array}{cccc}
 \frac{ 2\lambda(i \omega \sin(\omega t)+ \omega_p\cos(\omega t) )}{\omega^2-\omega_p^2} & -\frac{g}{\omega_p-\omega_q} \\
 -\frac{g}{\omega_p+\omega_q} & \frac{ 2\lambda(i\omega \sin(\omega t)+ \omega_p\cos(\omega t) )}{\omega^2-\omega_p^2}  
 \end{array} \right).
 \end{split}
\end{equation}
Substituting Eq. (\ref{eq:smatrix}) into Eq. (\ref{eq:zerothandsecond}), we arrive at the total effective Hamiltonian
\begin{equation}
\tilde{H}=\left( \begin{array}{cc}
\tilde{H}^{(0)}& 0\\ 
0 & \tilde{H}^{(1)}  \end{array} \right),
\end{equation}
where $\tilde{H}^{(0)}$ is the effective Hamiltonian in the lowest-energy subspace whose phonon number is zero and $\tilde{H}^{(1)}$ is the effective Hamiltonian of the subspace with one phonon.
The exact form of the effective Hamiltonian in the lowest subspace can be written as 
\begin{equation}
\begin{split}
\label{eq:effhamiltonian}
\tilde{H}^{(0)}=&H_0^{(0)}+\frac{1}{2}\left( S_1(t)H_1^{(10)}+H_1^{(01)}S_1^{\dagger}(t)\right)\\
=&-2\cos(\omega t)\frac{\lambda g \omega_p(\omega^2-2\omega^2_p+\omega_q^2) }{(\omega^2-\omega_p^2)(\omega_p^2-\omega_q^2)} \sigma_{x}\\
&+\omega_q \left(\frac{1}{2}-\frac{g ^2}{\omega_p^2-\omega_q^2} \right)\sigma_{z},
\end{split}
\end{equation}
where a two-dimensional identity matrix acting on the spin is omitted.

Now we consider the general system with $n$ phonons. Each two opposite spin states $|n \downarrow \rangle$ and $|n \uparrow \rangle$ with the same number of phonons form a subspace. Under the same precondition that $\Delta \geq \omega_p-\omega_q  \gg g,\lambda$, using the Schrieffer-Wolff transformation on the whole space, we find the effective Hamiltonian with $n$ phonons, 
\begin{equation}
\begin{split}
&\tilde{H}_{n}=  \tilde{H}_{n0}+ \tilde{H}_{n1}, \\
&\tilde{H}_{n0}=   \left(\frac{1}{2}-\frac{g ^2 (2n+1)}{\omega_p^2-\omega_q^2}\right)\omega_q\sigma_{z}+\omega_p n \\
&\tilde{H}_{n1}=  -\frac{2\lambda g \omega_p\cos(\omega t)(\omega^2-2\omega^2_p+\omega_q^2) }{(\omega^2-\omega_p^2)(\omega_p^2-\omega_q^2)} \sigma_{x}.
\end{split}
\label{eq:effwholehamiltonian}
\end{equation}
Here, the term including the identity matrix is eliminated already because it just produces a global phase. The coefficients of the second term with different numbers of phonons have different nonlinear coefficients. But the difference is small for small phonon numbers at low temperatures due to $\omega_p-\omega_q\gg g $ .

We transform the effective Hamiltonian into the Dirac picture with respect to $\tilde{H}_{n0}$ in Eq. (\ref{eq:effwholehamiltonian}). Assuming $\lambda\ll\omega_p$ and $\omega_ p+\omega \gg \Delta$, and applying a rotating wave approximation, the interaction part of the Hamiltonian becomes
\begin{equation}
\begin{split}
\tilde{H}_n^{}&=e^{i\tilde{H}_{n0}}\tilde{H}_{n1} e^{-i\tilde{H}_{n0}}\\
&=-\alpha (e^{2 i  \beta_n t} \sigma_{+}+ e^{-2 i   \beta_n t} \sigma_{-} ),
\end{split}
\label{eq:interaction}
\end{equation}
where
\begin{equation}
\begin{split}
&\alpha=\frac{\lambda g \omega_p(\omega^2-2\omega^2_p+\omega_q^2) }{(\omega^2-\omega_p^2)(\omega_p^2-\omega_q^2)}, \\ 
 &\beta_n=\frac{1}{2}\left(\omega_q\left(1-\frac{2(2 n +1) g ^2 }{\omega_p^2-\omega_q^2}\right)-\omega\right). 
 \end{split}
 \end{equation}
We transform the total Hamiltonian into the Dirac picture with respect to $\tilde{H}'_{n0}=\omega_p n + \beta_n\sigma_{z}$ as
\begin{equation}
\begin{split}
&\tilde{H}'_n=\tilde{H}_{np}'+\tilde{H}_{nq}'\\
&\tilde{H}_{np}'=\omega_p n,  \ \tilde{H}_{nq}'=\beta_n\sigma_{z}-\alpha\sigma_{x}.
\label{eq:effJC_hamiltonian}
\end{split}
\end{equation}
Arbitrary rotations in the $x$-$z$ plane can be generated by $\tilde{H}_{nq}'$.
 The coefficients of $\sigma_{x}$ and $\sigma_{z}$ are adjustable by varying the strength and the frequency of the driving. ESR can be started or stopped by turning the driving field on or off. Arbitrary-angle rotations about arbitrary axes can be constructed by combining these two rotations.

For a better understanding of approaching the arbitrary angle rotations, it is useful to derive the expression for the time-evolution operator on the spins. 
We rewrite $\tilde{H}'_{nq}=\vec{b}\cdot\vec{\sigma}$ in Eq.(\ref{eq:effJC_hamiltonian}) with the vector $\vec{b}=(-\alpha, 0, \beta_n)^T$. 
For $e^{-i \vec{b} \cdot \vec {\sigma} t}=\cos(|b|t)\mathbbm{1}-i \sin(|b|t) (\hat{b} \cdot \sigma)$, the time-evolution operator is given as
\begin{equation} 
\begin{split}
&R_{\vec{b}}(\theta)=e^{-i\tilde{H}'_{nq}t}=e^{-i \vec{b} \cdot \vec {\sigma} t} \\
=&\left( 
\begin{array}{cc}
\cos (\frac{\theta}{2})-i \frac{\beta_n}{\Omega} \sin(\frac{\theta}{2})& i \frac{\alpha}{\Omega} \sin( \frac{\theta}{2}) \\
i \frac{\alpha}{\Omega} \sin( \frac{\theta}{2}) & \cos (\frac{\theta}{2})+i \frac{\beta_n}{\Omega} \sin(\frac{\theta}{2})
\end{array} \right),
\label{eq:evolution}
\end{split}
\end{equation}
where $\Omega= \sqrt{\alpha^2 +\beta_n^2}$ and $\theta=2 \Omega t$. Rotations $R_{\vec{b}}(\theta)$ rotate the spins about the vector $\vec{b}$ by the angle $\theta $. First, arbitrary-angle rotations of single spins about arbitrary axes in the $x$-$z$ plane, such as the rotations relating to $X$-, $Z$-, phase and Hadamard gates, can be obtained by a single operation as $R=e^{i\gamma}R_{\hat{b}}(\theta) $ where $e^{i\gamma}$ is a global phase shift. Arbitrary angles $\theta$ can be achieved by adjusting the pulse time $t$. The axes  $\vec{b}=(-\alpha, 0, \beta_n )^T$ are in the $x$-$z$ plane and they are adjustable by varying the frequency and the strength of the electric driving field. Second, arbitrary-angle rotations about arbitrary axes out of the $x$-$z$ plane, such as rotation relating to $Y$ gate, can also be obtained by multiple steps of operations $R=e^{i\gamma}R_{z}(\theta_1)R_{x}(\theta_2)R_{z}(\theta_3)
 $ with 
appropriate $\gamma$ and angles of rotations $\theta_i(i=1,2,3) $. In this way, arbitrary-angle rotations of the single spin about 
arbitrary axes and arbitrary unitary single-qubit gates can be achieved \cite{Yale2013}.

\section{Numerical simulation}
\label{sec:numerics}

\begin{figure}[t]
 {\includegraphics[width=0.45\textwidth]{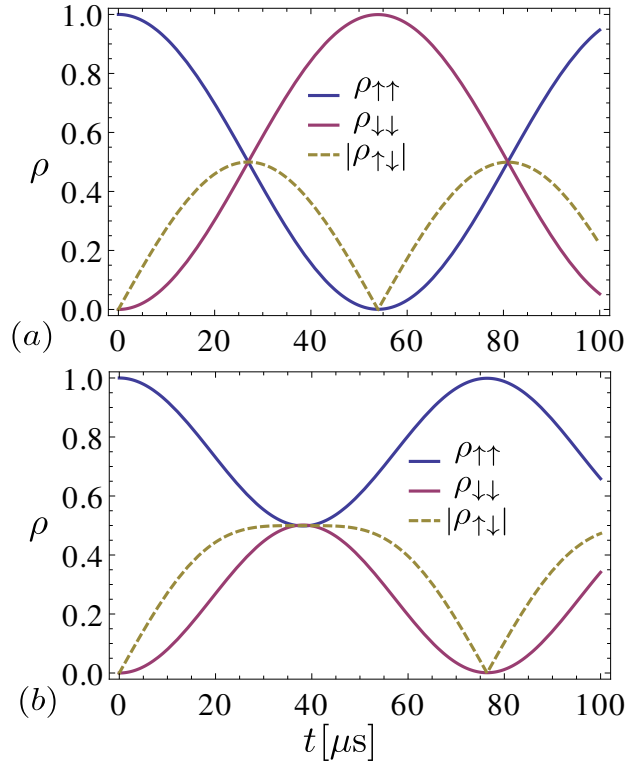}}
\caption{The time evolution of density matrix elements $\rho_{ij}=\langle i |\rho| j \rangle \ (i, j= \uparrow, \downarrow)$ under the conditions of (a) ESR with $\delta/2\pi=(\omega_q-\omega)/2\pi=0.003\ {\rm MHz}$ and (b) rotation about an axis in the  $x$-$z$ plane with $\delta/2\pi =0.012\ {\rm MHz}$ at zero temperature, with which one can achieve the Hadamard gate. The initial condition is $|0 \uparrow\rangle$. The blue (red) thick lines show the population of state $|0 \uparrow \rangle$ ($|0 \downarrow \rangle$). The dashed lines show the coherence terms between $|0 \uparrow \rangle$ and $|0 \downarrow \rangle$  in the ESR. The other parameters are $\lambda/2\pi = 0.8\ {\rm MHz}$, $\omega_q/2\pi =1.4\ {\rm GHz}$, $\omega_p/2\pi = 1.5\ {\rm GHz}$,  and $g/2\pi =0.56 \ {\rm MHz}$. 
\label{fig:Densitymatrixcompare}}
\end{figure}



At zero temperature, we can restrict our analysis to the states $|0 \uparrow \rangle$ and $|0 \downarrow \rangle$. Because the driving of the phonon states is suppressed by a large detuning, our system is weakly driven, and higher phonon states will not be involved in the time evolution. Therefore, there are only two states  $|0 \uparrow \rangle$ and $|0 \downarrow \rangle$ in the whole process at zero temperature. We simulate the time evolution for ESR of single electron spins and a rotation relating to the Hadamard gate at zero temperature in Fig. \ref{fig:Densitymatrixcompare}.

Now, we compose the $X$-, $Z$- and Hadamard gates at zero temperature. 
The $X$-gate is defined as 
\begin{equation} 
X=\sigma_x= \left( 
\begin{array}{cc}
0 & 1 \\
1 & 0 
\end{array} \right),
\end{equation}
and it can be achieved by setting $\alpha\neq0$ and $\beta_n=0$ in Eq.(\ref{eq:evolution}). To obtain $\beta_n=0$, we set $\omega=\omega_q(1-\frac{2 g ^2 }{\omega_p^2-\omega_q^2})$. 
We find $\omega=1399.997\ {\rm MHz}$  for the parameters $\lambda/2\pi = 0.8\ {\rm MHz}$,  $\omega_q/2\pi =1.4\ {\rm GHz}$, $\omega_p/2\pi = 1.5\ {\rm GHz}$, and $g/2\pi =0.56 \ {\rm MHz}$ (see Fig. \ref{fig:Densitymatrixcompare} (a)).
At the time points fulfilling $\sin(\theta/z)=1$ we can obtain the $X$ gate.

 The $Z$-gate is defined as
\begin{equation} 
Z=\sigma_z=\left( 
\begin{array}{cc}
1 & 0 \\
0 & -1 
\end{array} \right).
\end{equation}
  We can obtained the $Z$-gate by the off-resonance driving field with $\alpha=0$ and $\beta_n\neq0$ in Eq.(\ref{eq:evolution}). To achieve $\alpha=0$, the driving strength is set as $\lambda=0$. The parameters are chosen to be the same as for the X-gate. At the time points fulfilling $\sin(\theta/2)=1$ we can obtain the $Z$ gate.

The rotation of spin states for Hadamard gate,
\begin{equation} 
H= \frac{1}{\sqrt{2}}\left( 
\begin{array}{cc}
1 & 1 \\
1 & -1 
\end{array} \right),
\end{equation}
can be obtained by a rotation about an axis between $x$-and $z$-axes. By adjusting the frequency and the strength of the driving field, the axis of the rotation can be adjusted. The Hadamard gate is achieved by setting $\alpha=-\beta_n$, which requires $\omega/2\pi =1399.988 \ {\rm MHz}$, where we set $e^{i\gamma}=-i$ in $R=e^{i\gamma}R_{\hat{b}}(\theta) $. The parameters for the Hadamard gate are the same as those for the X-gate. A Hadamard gate is obtained at the time points fulfilling $\sin(\theta/2)=1$, for example at the contact point between $|0 \uparrow \rangle$ and $|0 \downarrow \rangle$ in Fig. \ref{fig:Densitymatrixcompare} (b).

\begin{figure}[t]
	\begin{center}
 {\includegraphics[width=0.45\textwidth]{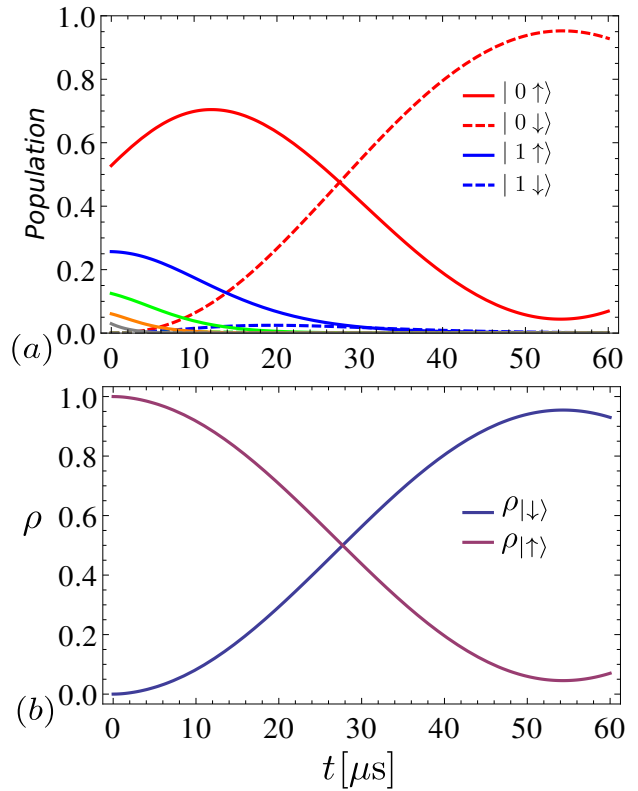}}
\caption { The time evolution of the population of spin states is shown at temperature 
$ T=100 \ {\rm mK}$. The initial state is $\rho_{\uparrow T}$. (a) The red thick (dashed) line shows the population of $|0 \uparrow \rangle$ ($|0 \downarrow \rangle$) and blue (dashed) line shows  $|1 \uparrow \rangle$ ($|1 \downarrow \rangle$). Green, orange and gray lines show $|2 \uparrow \rangle$, $|3 \uparrow \rangle$, and $|4 \uparrow \rangle$ separately, and the related states of spin down are damped out. (b) The total spin evolution $\rho_{\sigma}=\sum_n \rho_{n \sigma }$, $ \sigma= \uparrow ,\downarrow $ in thermal equilibrium. (Here we plot a figure with 
the 
maximum phonon 
number $n=4$.) The other parameters are $\lambda/2\pi = 0.8\ {\rm MHz}$, $\omega_p/2\pi = 1.5\ {\rm GHz}$, $\omega_q/2\pi=1.4\ {\rm GHz},\omega/2\pi=1399.997\ {\rm MHz}$ and $g/2\pi =0.56 \ {\rm MHz}$.
 \label{fig:finite}}
\end{center}
\end{figure}

Now we consider the case of a finite temperature. The distribution of phonons follows the Bose-Einstein statistics at finite temperature. The time evolution of ESR of all the spin states is simulated with a master equation at finite temperature (see Fig. \ref{fig:finite} (a)).  The initial state in thermal equilibrium has the form
\begin{equation}
\begin{split}
	\rho_{\sigma T}&= \frac{1}{Z} \sum_{n=0}^{\infty}e^{-n\hbar\omega_p / k_B T}|n \rangle \langle n | \otimes |\sigma\rangle \langle \sigma |\\
	&= \sum_{n=0}^{\infty} \rho_{n \sigma}|n \sigma \rangle \langle n \sigma|,
\end{split}
\end{equation}
where $\sigma=\uparrow,\downarrow$ and $Z=\sum_{n=0}^{\infty}e^{-n\hbar\omega_p / k_B T}$ is the partition function.
Because of the large detuning $\Delta$ and the vibrational damping at low temperature, the states with large numbers of phonons relax into the ground phonon state.
We now study the time evolution for ESR of the total spin states at finite temperature. The reduced density matrix of the spin is defined as $\rho_{\sigma }=\sum_n \rho_{n \sigma }$, $ \sigma= \uparrow ,\downarrow $. From the evolution of the total spin of states in Fig. \ref{fig:finite} (b), we can obtain ESR of the total spin states. Combining ESR with the rotations about the $z$-axis, arbitrary rotations on the Bloch sphere about arbitrary axes can be obtained at finite temperature.

\section{Analysis}
\label{sec:analysis}

Let us analyze the results in view of the envisioned aim of achieving arbitrary mechanically induced spin rotations. Different qubit states are connected by the driving and the spin-phonon coupling.
On one hand, the driving field couples to the phonon mode off-resonantly. A large frequency detuning $\Delta \gg g,\lambda$ is induced between the phonon mode and the driving field. The oscillator driving strength is quite small due to the large detuning.  Taking the state $|0\downarrow \rangle $ for example, it cannot be excited into the next higher phonon state $|1\downarrow\rangle$. 

\begin{figure}[t]
	\begin{center}
 {\includegraphics[width=0.4\textwidth]{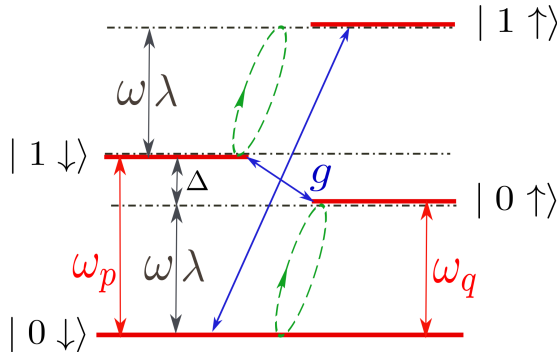}}
\caption { The energy level diagram of the combined spin-phonon states. The ac electric driving strength is denoted $\lambda$ and its frequency $\omega$.  The driving field couples the phonon mode with a large detuning of frequency $\Delta$.  The dash-dotted lines are the dressed states caused by the spin-phonon coupling. ESR  between two spin states with the same phonon numbers is obtained when $\omega$ approaches $\omega_q$ (green line).
 \label{fig:energylevels}}
\end{center}
\end{figure}

On the other hand, there is a large frequency difference between the phonon mode and the qubit mode, which is necessary for applying the Schrieffer-Wolff transformation. The combination of the spin-phonon coupling and related states results in the formation of dressed states (see Fig. \ref{fig:energylevels}). For example the coupling of $|1\downarrow \rangle$ to $|0 \uparrow \rangle$ with the strength $g$ results in two dressed states which are eigenstates of the non-driven ($\lambda=0$) Hamiltonian. Because of $\omega_p-\omega_q \gg g$, the two original states $|1 \downarrow \rangle$ and $|0 \uparrow \rangle$ are slightly mixed in the related dressed states. This means that one dressed state contains mainly either $|1 \downarrow \rangle$ or $|0 \uparrow \rangle$. 

 When the frequency of the driving approaches the frequency difference between the two dressed states which contain $|n \uparrow \rangle$ and $|n \downarrow \rangle$, then ESR between $|n \downarrow \rangle$ and  $|n \uparrow \rangle$ occurs. Due to $\Delta,\  \omega_p-\omega_q \gg g$, the driving effect for the phonons is limited and the spin rotations occur within the spin states with the same phonon numbers, which we can see from Eq.(\ref{eq:effJC_hamiltonian}). 

Our analytical theory is confirmed by numerical simulation.
From the two plots in Fig. \ref{fig:Densitymatrixcompare}, we see that ESR is sensitive to the frequency difference of $\delta=\omega_q-\omega $ and the efficiency of the transition of $|0 \downarrow \rangle \leftrightarrow |0 \uparrow \rangle$ changes obviously with a slight change of $\delta$. Perfect ESR with a full spin conversion is achieved only when the driving frequency $\omega$ approaches the frequency of the dressed states, which is shown to be at $\delta/2\pi=(\omega_q-\omega)/2\pi =0.003\ {\rm MHz}$ in Fig. \ref{fig:Densitymatrixcompare} (a).

A rotation about the $z$-axis can be obtained by applying an off-resonant
driving field. ESR is also adjustable by controlling the strength and the frequency of the external driving electric field. Hence, by using ESR and the rotation about $z$-axis together, we are able to obtain arbitrary-angle rotations of the spin about arbitrary axes through the one- or multiple-step operation.

\section{Conclusions}
\label{sec:conclusions}

In conclusion, the manipulation of a single electron spin in a suspended CNT using mechanical actuation at low temperature has been theoretically studied. We have proposed and analyzed the mechanical performance of rotations about the $x$-axis via ESR based on the curvature-induced spin-phonon coupling. The combination of ESR and rotations about the $z$-axis allows for arbitrary-angle rotations about arbitrary axes which are electrically controllable by varying the strength and the frequency of the external electric driving field. By choosing special pulse times, any single-qubit gates can be performed. We show that our proposal can be realized in experiment by numerical simulation with realistic parameters. Due to the long lifetime, spins can be detected during ESR. Importantly, the manipulation is all-electrical, thus it can be controlled and scaled up. Our proposal introduces a new way for manipulating spins with mechanical modes. Looking ahead, the fabrication and manipulation of entanglement between single spins in different QDs found on the same CNT or on adjacent CNTs as well as between the spin and the mechanical motion could be implemented by the  spin-phonon couplings, and two-qubit gates are expected to be achievable as well.

\section{Acknowledgments}

This work was supported by the DFG within the programs FOR 912 and SFB 767. Heng Wang acknowledges a scholarship from the State Scholarship Fund of China.


\bibliographystyle{apsrev}
\bibliography{Mech_Spin_Resonance-v8}

\end{document}